# From Learning Management System to Affective Tutoring system: a preliminary study

NADAUD Edouard
Learning, Data and Robotics Lab, ESIEA
Paris, Ile de France, France
edouard.nadaud@esiea.fr

GEOFFROY Thibault
Learning, Data and Robotics Lab, ESIEA
Paris, Ile de France, France
thibault.geoffroy@esiea.fr

KHELIFI Tesnim
Centre de Recherche en Informatique, Université Paris 1 Panthéon Sorbonne
Paris, Ile de France, France
Tesnim.Khelifi@etu.univ-paris1.fr

YAACOUB Antoun
Learning, Data and Robotics Lab, ESIEA
Paris, Ile de France, France
antoun.yaacoub@esiea.fr

Haidar Siba
Learning, Data and Robotics Lab, ESIEA
Paris, Ile de France, France
siba.haidar@esiea.fr

BEN RABAH Nourhène
Centre de Recherche en Informatique
Université Paris 1 Panthéon Sorbonne
Paris, Ile de France, France
Nourhene.ben-rabah@univ-paris1.fr

AUBIN Jean-Pierre
Learning, Data and Robotics Lab, ESIEA
Paris, Ile de France, France
jean-pierre.aubin@esiea.fr

PREVOST Lionel
Learning, Data and Robotics Lab, ESIEA
Paris, Ile de France, France
lionel.prevost@esiea.fr

LE GRAND Bénédicte
Centre de Recherche en Informatique, Université Paris 1 Panthéon Sorbonne
Paris, Ile de France, France
Benedicte.Le-Grand@univ-paris1.fr

## ABSTRACT

In this study, we investigate the combination of indicators, including performance, behavioral engagement, and emotional engagement, to identify students experiencing difficulties. We analyzed data from two primary sources: digital traces extracted from the Learning Management System (LMS) and images captured by students' webcams. The digital traces provided insights into students' interactions with the educational content, while the images were utilized to analyze their emotional expressions during learning activities. By utilizing real data collected from students at a French engineering school, recorded during the 2022-2023 academic year, we observed a correlation between positive emotional states and improved academic outcomes. These preliminary findings support the notion that emotions play a crucial role in differentiating between high-achieving and low-achieving students.

## KEYWORDS

Learning Management System, performance, behavioral engagement, emotional engagement, sentiment analysis,

## 1 INTRODUCTION

Over the past few years, higher education institutions have widely adopted online learning environments, notably Learning Management Systems (LMS), to maintain pedagogical continuity. Web-based platforms are now omnipresent in education, offering students and teachers easy access to- teaching resources, learning activities, and facilitating online exchanges. One of the major challenges facing these environments is student follow-up and the need for adequate support to prevent the risk of failure and encourage active participation [1]. Although LMSs offer functionalities for course organization, content delivery, and assessment, they can sometimes lack effective methods for ensuring individual supervision. Indeed, face-to-face interactions are replaced by online discussions and message exchanges, making it more difficult to spot potential problems [2]. To address this issue, several studies have proposed the use of Learning Analytics Dashboards [3] to monitor individual student performance. However, these dashboards primarily focus on academic performance neglecting other equally important indicators such as behavioral, emotional, and social engagement [4].

Therefore, the research question that the study aimed to address is: What is the impact of the combination of academic performance, emotional and behavioral engagement on the effectiveness of identifying students' difficulties? To address this question, we propose to analyze the connections between these indicators and to identify whether students' emotional engagement influences their academic performance and behavioral engagement. To achieve this, we present an innovative approach consisting of two phases. In the first phase, we collect raw data from two modalities: digital traces generated by the LMS and images captured by students' webcams. The digital traces encompass information about students' interactions with learning content, including time spent to, answers questions, and assessment results. Simultaneously, the images enable us to analyze students' emotional expressions. In the second phase, we establish the correlations between students' emotional engagement, academic performance, and behavioral engagement.

Our experiments confirm our hypothesis that positive emotional states can play a significant role in differentiating between high-achieving and low-achieving students.

The paper is organized as follows: section 2 outlines the main challenges. Section 3 conducts a comparative study of existing works. Section 4 presents the educational context along with the case study. Section 5 introduces the proposed methodology. Section 6 discusses the experimental results obtained in the study. Finally, section 7 presents conclusions and prospects.

## 2 MAJOR CHALLENGES

To develop our proposal, we encountered several challenges related to data collection and processing, requiring careful attention and innovative solutions:

**Technical**. One of the initial hurdles was the technical challenge, as not all students' laptops were equipped with webcams. This limitation restricted the total number of participants.

**General Data Protection Regulation (GDPR)**. We implemented careful planning and stringent measures to ensure data integrity and security. GDPR compliance meant that it was essential to put rigorous measures in place to protect participants' privacy. This involves obtaining informed consent from users, safeguarding sensitive information, restricting access to data, transferred on protected data servers, to authorized individuals only.

**Logistical**. We faced a logistical challenge regarding the need for sufficient storage space on our servers. The large amount of data generated, particularly in the form of images, required significant storage capacity.

**LMS database**. Understanding the structure of the Learning Management System (LMS) database, particularly Moodle, posed another challenge. Efficiently accessing and retrieving the data and metadata associated with the images necessitated an in-depth study of the Moodle system and data management.

**Matching captured facial images**. Matching a captured facial image to the specific question the student is answering at that moment added complexity. To address this, we implemented a sequential mode for tasks, enabling us to correlate each student's emotional state with the corresponding question.

**Data cleaning**. The final challenge involved data cleaning. We carefully filtered and eliminated images without faces, those with occlusions, lack of contrast, or those where the face detection algorithm failed to locate a face.

## 3 RELATED WORKS

In this section, we present five criteria for analyzing recent work proposing solutions to reduce the risk of failure, dropout, and support students in online learning environments. These criteria are learning indicators, identified sentiments, modality, objective, and data requirements. In this study, our focus was exclusively on empirical work presenting concrete practical cases. We made this choice because the existing literature often lacks specific proposals and remains predominantly theoretical [5]. So, we reviewed 10 recently published studies [8, 9

, 10, 11, 12, 13, 14, 15, 16, 17] (less than 5 years) and compared them on the following criteria.

**Learning Indicators**. Some studies focus on the performance indicator, as evident in studies [8], [12], [14] and [15]. Other studies also address the engagement indicator, such as [16] and [17], specifically focusing on emotional engagement in studies [9] and [10], and cognitive engagement in study [9]. Study [11] addresses both indicators, i.e., performance and engagement.

**Identified sentiments**. It is worth noting that some studies employ a polarity approach to identify feelings, categorizing them as positive, negative, or neutral. This approach is seen in works [10], [12], [13] and [14]. Other studies adopt a dimensional approach, relying on valence in sentiment analysis, as observed in study [15]. Some works combine the polarity dimension with the valence dimension, as illustrated in study [8]. Additionally, study [16] combines the arousal and polarity dimensions in sentiment analysis. However, certain studies opt for a categorical approach, identifying specific feelings, as described in [10] and [17].

**Modality**. We can observe that among the 10 analyzed studies, a common characteristic is that most of them rely on unimodal data sources, predominantly derived from the e-learning platform. However, study [14] stands out for its multimodal approach, where researchers collected data from both the university database and a mobile application that generated data based on student activity.

**Objective**. The reviewed studies encompass a range of objectives. The most common objective is the prediction of students' academic performance, as seen in studies [16], [11], [14] and [15]. Some works focus on predicting engagement, as in [9] and [17], or predicting the risk of dropping out, as in [14]. Other studies aim to provide personalized interventions based on students' difficulties, such as [9], which suggests a personalized teaching strategy by adapting the platform to students with engagement problems, similar to studies [16] and [17]. Additionally, study [10] adapts exercises based on detected emotions. Other studies offer personalized feedback to students, as in study [10], or utilize student feedback to improve the learning process, as in [13].

**Data requirements**. The utilization of educational data is subject to various requirements. We have defined five requirements, inspired by those defined by the European Commission in the context of trustworthy artificial intelligence [7].

(R1) Data confidentiality.

(R2) Informed consent.

(R3) Data anonymization.

(R4) Transparency.

(R5) Diversity.

Table 1 presents the final criterion for comparison between studies, which is Data requirements. It is noteworthy that all the studies implemented data anonymization measures (R3). Considering the other requirements, most studies unfortunately lack rigor, probably because confidentiality rules are not the same all around the world.

**TABLE 1. Comparative table of data requirements.**

| Data Requirements across reviewed studies | | | | |
|---|---|---|---|---|
| R1 | R2 | R3 | R4 | R5 |
| 1 | 1 | 10 | 0 | 2 |

## 4 EDUCATIONAL CONTEXT AND CASE STUDY

Our study utilized authentic data from a French engineering school. During our study, the school, and its research lab, adhered to a rigorous data collection methodology that respected individual rights and complied with personal data protection regulations, including GDPR.

We obtained informed consent from each participant, ensuring that students' rights were respected, and their data remained confidential. Participants were informed about the purpose of data usage and their rights regarding their data, including the right to contest, access, correct, delete, and refuse data usage for purposes other than research. To safeguard anonymity, all information that could potentially identify participants was removed from the data.

To promote equity and equal opportunities, we ensured the use of diverse data in terms of gender, aiming to eliminate gender stereotypes and adapt pedagogical interventions to all learners regardless of their gender. We also made the results of our experiments transparent and comprehensible to all participants to establish trust.

By adhering to ethical and data protection principles, we ensured ethical research practices while delving into the relationship between emotions and learning.

The research was conducted within the context of the Machine Learning course, which is taught to second-year engineering students, equivalent to the fourth year of higher education. The assessment system for the course included three quizzes, a practical report, and a final exam. We employed a multimodal approach to collect educational data, utilizing two primary sources: (a) digital traces generated by the Learning Management System (LMS) to analyze the performance and engagement of 324 students, and (b) facial expression images recorded via cameras capturing the expressions of 89 students.

## 5 PROPOSED METHODOLOGY

Our approach consists of two key phases: the collection of pertinent data and the analysis of this data to provide support to students in an online course.

### 5.1 Performance and behavioral engagement

Learning Analytics based on LMS [27] in blended or online learning environments relies on digital traces of learning that describe students' activity and engagement. They can be directly exploited or used based on indicators that describe behavioral engagement, affective engagement, performance, and other relevant factors. In this context, we present in this section two key indicators [18] performance and engagement, which measures active participation, degree of involvement, and motivation of students in the learning process [19].

*5.1.1 Performance indicator*
The performance indicator is used to give an estimation of a student's level of success in the e-learning module by calculating the average of his/her grades in the executable activities. All grades are assigned the default coefficient 1 to calculate the average. We also computed a particular measure: the Discriminative Grade (DG) per question per student defined as follows:

$$DG = \text{Grade on the question} \times DE \quad (1)$$

Where DE is the Discrimination Efficiency of a quiz question. It is a statistical measure computed by the LMS which quantifies how well the question can differentiate between high and low-performing students.

*5.1.2 Behavioral engagement indicators*
Concentration index (CI) is computed using the following equation, proposed in [20].

$$CI = (Emotion\_Weight \times Gaze\_Weight) / \sum Emotion\_Weight$$

*Emotion_weight* depends on the basic (Ekman's) emotion detected in each frame, while *Gaze_weight* depends on the degree of eye opening and gaze orientation. The product is normalized to vary between 0 and 1. The value of CI is thresholded to evaluate the student Engagement Degree (ED) taking three ordinal values, from disengaged to highly engaged [21].
To compute this indicator, we used the python dlib library and a pre-trained 21-layered neural network dedicated to emotion recognition trained on the FER'13 dataset.
We also use the Time Spent (TS) on each question is directly extracted from Moodle.

### 5.2 Emotional engagement

We also analyzed the facial expressions of students who participated voluntarily while they were focused on their graded work. Out of 89 students, only 56 (3 women and 53 men) had usable facial expression data due to technical issues (detailed in section 2). The students answered quiz questions sequentially, and an image was taken every 5 seconds. To evaluate the emotional experiences, we used Russell's dimensional model and particularly valence [30], that quantifies the degree of emotion positiveness expressed by students during the quiz interactions. We used the model described in [22], that was trained on AffectNet and get a RMSE of 0.24.

## 6 EXPERIMENTAL RESULTS

### 6.1 Performance and behavioral engagement

In this section, we review the results of the global engagement and performance indicators. We also present a detailed view of the obtained results. The performance of the class was measured and found to be 54%. Simultaneously, we observed a low engagement score of 27%. Based on these findings, we can conclude that the students' low level of engagement had an impact on their academic results, as indicated by the performance score, which did not reach a satisfactory level.

In our study, engagement is measured by the indicator CI. Performance, on the other hand, is measured by the average of the discriminative grade and the grade attributed to each question. Figure 2 illustrates a bar chart depicting the performance progression of student number 3010 across the seven questions of quiz number 2. It is evident from the chart that there exists a direct correlation between performance and engagement. Specifically, the greater the level of investment exhibited by a student in answering a question, the higher their performance tends to be.

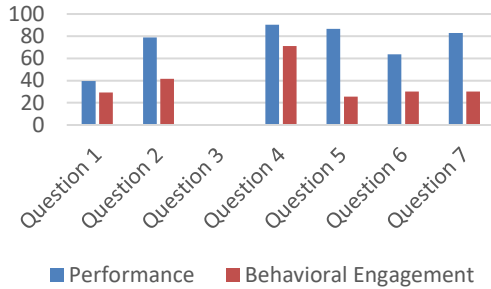

**Figure 2. Performance and behavioral engagement scores for student 3010 on Quiz 2.**

## 6.1 Connection between performance, behavioral engagement, and emotional state

In this section, we analyzed the valence indicator (VAL) for each image captured during student number 3010's response to the 7 questions in quiz 2. For each question, we recorded the score, the discrimination efficiency (DE), the discriminant score (DS), the time spent (TS), the most significant image (MSI) which presents the image with the highest absolute valence score captured during the response to each question corresponding to the student's emotional state at the time of image capture, and finally the valence score (VAL). Table II presents these scores.

Looking at table II, we can see for example that during the sixth question (Q6), the student received a grade of 50%, and the question had a DE of 33,66. Therefore, the DS was computed as 16,83. The valence score of -0.16, indicating a slightly negative emotional state. This process was repeated for each question in the quiz. As seen from the table II, there is a variety of grades, DE, DS, TS and valence scores across the seven questions, highlighting the dynamic nature of the learning process and the varying emotional states experienced by the student.

**Table II. Representative data for student number 3859 across seven questions of quiz number 2.**

| Q | Grade | DE | DG | TS(s) | VAL | ED |
|---|-------|-------|-------|-------|-------|------|
| 1 | 0 | 71,64 | 0 | 99 | -0,23 | low |
| 2 | 100 | 74,49 | 74,49 | 27 | -0,31 | low |
| 3 | 0 | 55,76 | 0 | 46 | -0,20 | nice |
| 4 | 100 | 79,79 | 79,79 | 21 | -0,28 | nice |
| 5 | 100 | 45,56 | 45,56 | 39 | -0,33 | nice |
| 6 | 50 | 33,66 | 16,83 | 519 | -0,16 | low |
| 7 | 100 | 45,64 | 45,64 | 52 | -0,10 | low |

We now analyze the possible links between valence and engagement. For this we have divided our valence indicator into 10 categories to perform a chi-square analysis with the engagement degree. We check the dependence at ☐=5% significance level between these two variables on the whole dataset and show that these to variables were statistically dependent.

Therefore, we propose that positive emotions can contribute to distinguishing between different levels of student performance. Lastly, a negative correlation is observed between (VAL) and (TS).

## 6.2 Discussion about collected data

We display several images in figure 2 to show that collected images are more "in the wild" images than "in constrained environments" (see discussion in section 2). This fact can explain the relative performances of emotion and affect estimator. Moreover, some students exhibited focused and attentive expressions, characterized by concentrated gazes and relaxed facial features, indicating a positive emotional state associated with deep engagement. On the other hand, certain images portrayed students with furrowed brows, tense postures, or distracted looks, suggesting negative emotions such as confusion or lack of interest. This indicates that (1) valence is not sufficient to characterize affect state and that we need to detect other "affect events" and (2) the sampling period (5sec) is probably too high to detect such event

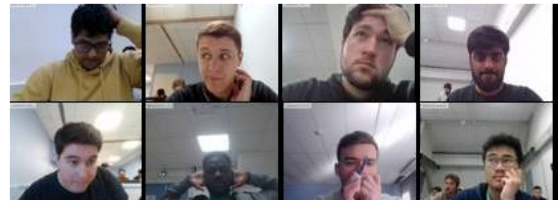

**Figure 2. Wide spectrum of affective states**

## 7 CONCLUSION AND FUTURE WORKS

In this paper, we have presented an approach to investigate the relationships between academic performance, behavioral engagement, and emotional engagement with the goal of identifying struggling students. Throughout the study, we encountered various challenges related to data collection and analysis, including issues of confidentiality, ethics, data cleaning, and matching captured facial images, among others. The experimental findings have demonstrated the significant role played by positive emotional states in distinguishing between high-achieving and low-achieving students.

However, it is important to acknowledge that this study has certain limitations that should be taken into consideration. Firstly, the aspect of emotion recognition was limited to students with a webcam, which may restrict the generalizability of the results to a broader student population. Additionally, the study had to comply with GDPR regulations and ensure data integrity, which added complexity to the research process.

In future work, we plan to expand our data collection efforts to gather more comprehensive datasets. We also intend to explore other dimensions, such as arousal for emotional engagement and social engagement, to enhance our understanding of the factors that influence student success. By doing so, we aim to develop more precise strategies for identifying and supporting struggling students.